\documentclass[%
 reprint,
superscriptaddress,
amsmath,amssymb,
aps,
pra,
longbibliography,
]{revtex4-2}

\usepackage{graphicx}
\usepackage{dcolumn}
\usepackage{bm}
\usepackage{bbm}
\usepackage{amsfonts}
\usepackage{amsthm}
\usepackage{tikz}
\usepackage{bm}
\usepackage{physics}
\usepackage{comment}
\usepackage{xcolor}
\usepackage{blindtext}
\usepackage{pgfplots,pgfplotstable}
\usepgfplotslibrary{polar}
\usepgfplotslibrary{colormaps}
\pgfplotsset{compat=newest}
\usepackage{hyperref}

\definecolor{DTgreen}{rgb}{.1, .5, .2}

\definecolor{HS}{rgb}{0.0,0.5,0.0}
\definecolor{SN}{rgb}{1, 0.5, 0.31}
\definecolor{SSN}{rgb}{0.87, 1.0, 0.0}
\definecolor{OSN}{rgb}{1.0, 0.44, 0.37}
\newcommand{\re}[1]{\Re[#1]}
\newcommand{\im}[1]{\Im[#1]}
\providecommand{\R}{\mathbb{R}}

\renewcommand{\O}{O}

\newcommand{\vac}{\ket{\mathrm{vac}}}
\providecommand{\abs}[1]{\left\lvert#1\right\rvert}

\renewcommand{\d}{\mathrm{d}}
\newcommand{\e}{\mathrm{e}}
\newcommand{\ii}{\mathrm{i}}
\newcommand{\diag}{\mathrm{diag}}

\newcommand{\NS}{N_\mathrm{S}}
\newcommand{\ND}{N_\mathrm{D}}

\newcommand{\Uphi}{U_{\varphi}}
\newcommand{\hUphi}{\hat{U}_{\varphi}}

\newcommand{\hx}[1]{\hat{x}_{#1,\theta_{#1}}}
\newcommand{\vx}{\boldsymbol{x}}

\newcommand{\Var}{\mathrm{Var}}
\renewcommand{\Vec}[1]{\boldsymbol{#1}}

\theoremstyle{theorem}

\theoremstyle{definition}

\begin{document}


\title{Non-adaptive Heisenberg-limited metrology with multi-channel homodyne measurements}

\author{Danilo Triggiani}
\email{danilo.triggiani@port.ac.uk}
\affiliation{School of Mathematics and Physics, University of Portsmouth, Portsmouth PO1 3QL, UK}

\author{Paolo Facchi}
\affiliation{Dipartimento di Fisica and MECENAS, Università di Bari, I-70126 Bari, Italy }
\affiliation{INFN, Sezione di Bari, I-70126 Bari, Italy}

\author{Vincenzo Tamma}
\email{vincenzo.tamma@port.ac.uk}
\affiliation{School of Mathematics and Physics, University of Portsmouth, Portsmouth PO1 3QL, UK}
\affiliation{Institute of Cosmology and Gravitation, University of Portsmouth, Portsmouth PO1 3FX, UK}





\date{\today}

\begin{abstract}
We show a protocol achieving the ultimate Heisenberg-scaling sensitivity in the estimation of a parameter encoded in a generic linear network, without employing any auxiliary networks, and without the need of any prior information on the parameter nor on the network structure. As a result, this protocol does not require a prior coarse estimation of the parameter, nor an adaptation of the network. The scheme we analyse consists of a single-mode squeezed state and homodyne detectors in each of the $M$ output channels of the network encoding the parameter, making it feasible for experimental applications.
\end{abstract}

\pacs{Valid PACS appear here}
\maketitle


\section{Introduction}

Increasing the level of precision achievable in the estimation of physical properties of systems, such as temperatures, optical lengths and magnitude of external fields among others, is one of the multiple applications of quantum technologies that have been extensively studied in the recent years. 
In particular, the goal of quantum metrology --- the field of science laying between quantum mechanics and estimation theory --- is to propose and analyse estimation protocols that surpass the precision achievable by classical strategies by employing quantum probes and quantum measurement schemes. 
In fact it is well known that the classical limit on the precision achievable in the estimation of an unknown parameter when employing $N$ probes, known as shot-noise limit, for which the error is of order $1/\sqrt{N}$, can be surpassed by quantum strategies achieving the ultimate Heisenberg limit, where the estimation scales as $1/N$~\cite{Giovannetti2004, Giovannetti2006, Dowling2008, Giovannetti2011,Dowling2015, DePasquale2015, Zhou2018, Ge2018,Qian2019}

The first proposed protocols reaching Heisenberg scaling sensitivity heavily employed entanglement as a metrological resource~\cite{Giovannetti2004, Giovannetti2006, Dowling2008}, and several entanglement-based strategies have been recently studied with interesting results especially in the cases of simultaneous estimation of multiple parameters with non-commuting generators~\cite{Dowling2015, DePasquale2015, Proctor2017, Zhou2018, Ge2018,Qian2019}. 
Nonetheless, the entanglement fragility and the complicated procedures needed to generate entangled metrological probes, such as NOON or GHZ states~\cite{Greenberger1989,Kok2002}, are two of the challenges that stimulated the search for more feasible estimation schemes making use of protocols implementing metrological resources that are easier to generate and to manipulate. 
Squeezed light~\cite{scully1997,schleich2011} manifests useful properties (e.g. robustness to decoherence, relatively easy implementation, reduced noise below the vacuum shot-noise) which make it a perfect candidate as a feasible metrological resource~\cite{Monras2006, Pezze2008, Lang2013, Ligo2013, Maccone2020}. 
Motivated by these favourable properties, many works have recently focused on the analysis and proposal of Gaussian metrological schemes, namely involving squeezed states as probes and homodyne detection as measurement, and the ultimate precision that these can achieve in the estimation of a single localised parameter~\cite{Monras2006, Pezze2008, Aspachs2009, Lang2013, Ligo2013, Oh2019}, a function of parameters~\cite{Gatto2020,Xia2020,Triggiani2021}, or a single distributed parameter, such as the temperature or the electromagnetic field, affecting several component of the network~\cite{Zhuang2018, Ge2018, Matsubara2019, Gatto2019, Guo2020, Gramegna2021njp, Gramegna2021Typicality}. 
Interestingly, it has been recently found that it is always possible to reach Heisenberg scaling sensitivity regardless of the structure of the network encoding the parameter, only employing a single squeezed-vacuum state, a single homodyne measurement, and an auxiliary network suitably engineered, whose preparation only requires a knowledge on the unknown parameter that can be obtained by a classical measurement~\cite{Matsubara2019,Gramegna2021njp,Gramegna2021Typicality}. 
The need for an auxiliary stage in such protocols arises from the fact that in general the probe is scattered by the network in all its output ports, while only a single port is eventually measured through homodyne detection, so that an auxiliary network is required in order to refocus the probe on the only channel observed. 
A question that naturally arises is whether incrementing the number of observed channels would ease, if not completely lift, the requirement of an auxiliary stage and ultimately the requirement of a prior classical knowledge on the unknown parameter. 
Moreover, different Gaussian protocols rely on encoding the information about the unknown parameter on the displacement of the probe, requiring that a portion of the resources in the probe are employed in a non-vanishing displacement~\cite{Guo2020,Grace2021}. 
Despite concentrating all the photons in the squeezing is known to be the optimal allocation of the resources in the probe~\cite{Matsubara2019}, encoding the parameter into a non-vanishing displacement can reduce the estimation process into the relatively simple task of inferring the parameter from the expectation value of a Gaussian probability density function~\cite{Guo2020}.

In this work we investigate the ultimate precision achievable in the estimation of a parameter encoded in a generic linear network, when employing a single-mode squeezed coherent Gaussian state and performing homodyne detection on all the output channels. 
We show that, without making any assumption on the structure of the linear network nor on the nature of the parameter, it is always possible to reach Heisenberg scaling sensitivity with such setup, without the use of any auxiliary network. 
This allows for estimation protocols not requiring a preparatory stage nor a prior coarse estimation of the parameter, as opposed to single-channel homodyne protocols.
We also show that two independent contributions on the precision arise from our analysis: one originated from the presence of displaced photons in addition to squeezed photons, the other from the squeezing of the probe. 
We find that both contributions can reach Heisenberg scaling sensitivity independently, and this can be achieved expectedly when the local oscillators phases are chosen such that the noise in the outcome is reduced, namely when the squeezed quadratures are observed in each output channel.
Although it is not required to reach the Heisenberg scaling sensitivity, the presence of an auxiliary network in general affect the precision of the estimation through a pre-factor multiplying the scaling.
This comes in useful in those cases were priority is given to increasing the  precision, at the expenses of engineering an auxiliary network to be added before the estimation protocol is started.

\section{Setup}

\begin{figure}[t]
\centering
\includegraphics[width=.95\columnwidth]{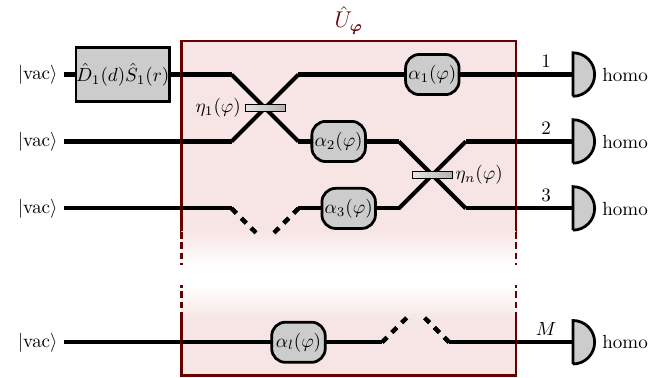}
\caption{Optical setup for the estimation at the Heisenberg-scaling precision of a single unknown parameter $\varphi$ encoded arbitrarily in a generic passive $M$-channel linear network. The parameter can be either localised in a single element of the network, or represent a global property affecting several components, such as a temperature or a magnetic field. A single source of coherent squeezed states with $N = \ND + \NS$ average photons, where $\ND=d^2$ and $\NS=\sinh^2 r$ are the number of displaced and squeezed photons respectively, is employed in a single input channel (the first in figure), while homodyne measurements are performed in every output channel. }
\label{fig:GenericSetup}
\end{figure}

Let us consider a generic $M\times M$ passive linear network whose action on any injected photon probe is given by the unitary operator $\hUphi$, in which the unknown parameter $\varphi$ to be estimated is encoded in an arbitrary manner. 
The linearity and passivity of the network allow us to describe it with an $M\times M$ unitary \emph{matrix} $\Uphi$ related to the evolution operator $\hUphi$ by
\begin{equation}
\hUphi^\dag \hat{a}_i \hUphi = \sum_{j=1}^M (\Uphi)_{ij}\hat{a}_j.
\end{equation}
The input probe is prepared in a single-mode squeezed coherent state $\ket{\Psi_{\mathrm{in}}} =\hat{D}_1(d)\hat{S}_1(r)\vac$ with $N=\sinh^2 r + d^2 \equiv \NS + \ND$ average number photons, where $\hat{S}_1(r) = \exp \bigl(r(\hat{a}_1^{\dag2}-\hat{a}_1^{2})/2\bigr)$ is the squeezing operator with real squeezing parameter $r$, and $\hat{D}_1(d) = \exp\bigl(d(\hat{a}^\dag_1-\hat{a}_1)/\sqrt{2}\bigr)$ is the displacement operator with real displacement $d$, and it is injected in one input channel, say the first, of the linear network.
In this case, only the first row of $\Uphi$ is relevant in this protocol
\begin{equation}
(\Uphi)_{1j} = \sqrt{P_j}\e^{\ii \bar{\gamma}_j},
\end{equation}
where we have made explicit the probability $P_j$ that each photon exits from the $j$-th output port of the network, and the phase $\bar{\gamma}_j$ acquired in the process, with $j$ from $1$ to $M$. The unitarity of $\Uphi$ assures that $\sum_j P_j = 1$.

A homodyne detection is then performed at each of the output channels, and the quadratures $\hx{i}$ are measured, where $\theta_i$ is the $i$-th local oscillator reference phase, from which we want to infer the value of $\varphi$.
Due to the Gaussian nature of the scheme, the joint probability distribution $p(\vx|\varphi)$ associated with the $M$-mode homodyne measurement is Gaussian
\begin{equation}
\label{eq:pdf}
p(\vx|\varphi) = \frac{1}{\sqrt{(2\pi)^M \abs{\Sigma}}}\exp\Bigl[-\frac{(\vx-\Vec{\mu})^\mathrm{T}\Sigma^{-1}(\vx-\Vec{\mu})}{2}\Bigr].
\end{equation}
Here, $\Sigma$ is the $M\times M$ covariance matrix with elements (see appendix~\ref{app:Prob})
\begin{eqnarray}
\Sigma_{ij} &=& \frac{\delta_{ij}}{2} + \sqrt{P_i P_j}\bigl(\cos(\gamma_i-\gamma_j)\sinh(r)^2
\nonumber\\
& & \qquad \qquad \qquad +\cos(\gamma_i+\gamma_j)\cosh(r)\sinh(r)\bigr),\quad 
\label{eq:CovMat}
\end{eqnarray}
where $\delta_{ij}$ is the Kronecker delta, $\gamma_i=\bar{\gamma}_i-\theta_i$ is the phase delay at the output of the $i$-th channel relative to the correspondent local oscillator, and $\abs{\Sigma}$ is the determinant of $\Sigma$, which   reads (see appendix~\ref{app:Prob})
\begin{align}
\abs{\Sigma} = \frac{1}{2^M} &+ \frac{\sinh(r)}{2^{M-1}}\sum\limits_{i=1}^M P_i (\sinh(r) + \cos(2\gamma_i)\cosh(r)) \notag\\
&-\frac{\sinh^2(r)}{2^{M-2}}\sum\limits_{i = 1}^M \sum\limits_{j = i+1}^M P_i P_j \sin^2(\gamma_i-\gamma_j),
\label{eq:Det}
\end{align}
and $\Vec{\mu}$ is the mean vector
\begin{equation}
\mu_i = d\sqrt{P_i}\cos\gamma_i.
\label{eq:AverageVector}
\end{equation}

For any given unbiased estimator $\tilde{\varphi}$, the statistical error in the estimation of $\varphi$ after $\nu$ iterations of the measurement is limited by the Cramér-Rao bound (CRB)~\cite{cramer1999mathematical}
\begin{equation}
\Var[\tilde{\varphi}] \geq \frac{1}{\nu \mathcal{F}(\varphi)},
\label{eq:CRb}
\end{equation}
where $\mathcal{F}(\varphi)$ is the Fisher Information
\begin{equation}
\mathcal{F}(\varphi) = \int \d\vx\ p(\vx|\varphi)\bigl(\partial_\varphi\log p(\vx|\varphi)\bigr)^2,
\label{eq:FisherDef}
\end{equation}
associated with the Gaussian  distribution~\eqref{eq:pdf}, and reads (see appendix~\ref{app:Fisher})
\begin{align}
\mathcal{F}(\varphi) &= \frac{1}{\abs{\Sigma}}\partial_\varphi\Vec{\mu}^\mathrm{T}C\partial_\varphi \Vec{\mu} +  \frac{1}{2}\left(\frac{\partial_\varphi\abs{\Sigma}}{\abs{\Sigma}}\right)^2 \notag\\
&\quad - \frac{1}{2\abs{\Sigma}}\Tr[(\partial_\varphi\Sigma)(\partial_\varphi C)],
\label{eq:Fisher}
\end{align}
where $C=\abs{\Sigma}\Sigma^{-1}$ is the cofactor matrix of $\Sigma$ and $\Tr[\cdot]$ denotes the trace. 
In the following we will discuss in detail expression~\eqref{eq:Fisher} in the asymptotic limit of large $N$, showing which condition must be met in order for this setup to reach Heisenberg-scaling precision in the estimation of $\varphi$, and compare differences and advantages with respect to the Heisenberg-scaling single-homodyne schemes~\cite{Gramegna2021njp,Gramegna2021Typicality}. 

We conclude this section by remarking that, in the case of a single channel, $M=1$, the last term in the right-hand side of~\eqref{eq:Fisher} vanishes, and thus the only relevant terms are the first two, containing the derivative of the mean $\Vec{\mu}$ and of the determinant of $\Sigma$, which reduces to the variance of a single-homodyne measurement.
Interestingly enough, we will show that also in the multi-homodyne case, only the first two terms are relevant for the Heisenberg scaling in the asymptotic regime.

\section{Heisenberg scaling of the Fisher Information }
\label{sec:Fisher}

In order to investigate the asymptotic behaviour of the Fisher Information~\eqref{eq:Fisher}, it is convenient to express the elements of the cofactor matrix $C$ in terms of the squeezing factor $r$ (see appendix~\ref{app:Fisher}):
\begin{align}
C_{ss} &= \frac{1}{2^{M-1}} + \frac{1}{2^{M-2}}\sum\limits_{\substack{i=1 \\ i \neq s}}^M (\Sigma_{ii} - \frac{1}{2}) - \frac{1}{2^{M-3}}\sum\limits_{\substack{i=1 \\ i \neq s}}^M\sum\limits_{\substack{j=i+1 \\ j \neq s}}^M S_{iij},\notag\\
C_{st} &= -\frac{1}{2^{M-2}}\Sigma_{st} + \frac{1}{2^{M-3}}\sum\limits_{\substack{i=1 \\ i\neq s,t}}^M S_{sti}\ ,\quad s\neq t,
\label{eq:CofMat}
\end{align}
where 
\begin{equation}
S_{sti} = \sinh^2{r}\sqrt{P_s P_t} P_i \sin(\gamma_s - \gamma_i)\sin(\gamma_t-\gamma_i).
\label{eq:STerms}
\end{equation}

Notice that every element of $C$ in the previous expressions, and of $\Sigma$ in~\eqref{eq:CovMat}, scale at most as quick as $N$, namely $C_{st}=\O(\NS)$ and $\Sigma_{st}=\O(\NS)$, while the mean vector $\Vec{\mu}$ is of order $\O(\sqrt{\ND})$ (see Appendix~\ref{app:Asymptotics}), and the same asymptotic bounds hold for their derivatives with respect to $\varphi$, since neither $P_i$ nor $\bar{\gamma}_i$ depend on $N$. 
For this reason, in order for the Fisher information in~\eqref{eq:Fisher} to asymptotically grow with Heisenberg scaling, it is essential to study the asymptotics of the determinant $\abs{\Sigma}$ in~\eqref{eq:Det}, and find the conditions for which it does not grow with $N$.

In fact, it is evident from equation~\eqref{eq:Det} that in general $\abs{\Sigma}=\O(\NS)$, and we show in appendix~\ref{app:Asymptotics} that the necessary condition for it to scale slower than $\NS$ is that the relative phases $\gamma_i$ tend to $\pm \pi/2$ for large $\NS$: in other words, the larger the number of photons employed in the squeezing of the probe to reach higher precisions, the closer the local oscillator phase needs to be tuned to  the minimum-variance quadrature of each mode. 

More precisely, as shown in Appendix~\ref{app:Asymptotics}, the conditions to reach Heisenberg scaling in the Fisher information~\eqref{eq:Fisher}, read
\begin{equation}
\gamma_i = \pm\frac{\pi}{2} + \O(N_S^{-1}),\qquad i=1,\dots, M.
\label{eq:HLConditions}
\end{equation}
When these conditions hold, we can introduce the finite quantities $k_i =\lim_{\NS\rightarrow\infty} \NS(\gamma_i\mp\pi/2)$, and the determinant $\abs{\Sigma}$ reduces to 
\begin{equation}
\abs{\Sigma} = \frac{1}{2^{M-2}\NS}\biggl(\Bigl(\sum\limits_{i=1}^M P_i k_i\Bigr)^2 +\frac{1}{16}\biggr),
\label{eq:DetAsymptotic}
\end{equation}
while $\partial_\varphi \abs{\Sigma}$, $\partial_\varphi \Sigma$, $\partial_\varphi C$ and $C$ tend to constant values, and $\partial_\varphi \Vec{\mu}$ scales as $\sqrt{\ND}$, thus making only the first two terms of the Fisher information dominant for large $N$.

As expected, the determinant of the covariance matrix $\Sigma$ reaches its minimum value when $\gamma_i = \pi/2$, or $k_i=0$ for $i=1,\dots,M$, namely when the squeezed quadratures are measured. When conditions~\eqref{eq:HLConditions} are met, we can neglect the trace term in equation~\eqref{eq:Fisher}, and we can write 
\begin{align}
\mathcal{F}(\varphi) &\simeq \frac{1}{\abs{\Sigma}}\partial_\varphi\Vec{\mu}^\mathrm{T}C\partial_\varphi \Vec{\mu} + \frac{1}{2}\left(\frac{\partial_\varphi\abs{\Sigma}}{\abs{\Sigma}}\right)^2 \notag\\
&\simeq 8(\partial\gamma)_\mathrm{avg}^2\left( \zeta(k_\mathrm{avg}) N_D N_S 
+ \varrho\left(k_\mathrm{avg}\right) N_S^2\right) ,
\label{eq:FisherAsymptotic}
\end{align}
where $k_\mathrm{avg}\equiv\sum\limits_{i=1}^M P_i k_i$, $(\partial\gamma)_\mathrm{avg} \equiv \sum\limits_{i=1}^M P_i \partial_\varphi\gamma_i$, and $\varrho(x) = (8x)^2/(16x^2 + 1)^2$ and $\zeta(x)=(16x^2 + 1)^{-1}$ are positive, even function which reach their maxima at $x=\pm 1/4$ and $x=0$, respectively, namely $\varrho(1/4)=1$ and $\zeta(0)=1$. 
The Cramér-Rao bound~\eqref{eq:CRb} with the Fisher information~\eqref{eq:FisherAsymptotic} is saturated for large $\nu$ by the maximum-likelihood estimator (shown in appendix~\ref{app:MLE}), and thus Heisenberg-scaling precision can be achieved.

Noticeably, both terms in the asymptotic Fisher Information~\eqref{eq:FisherAsymptotic} give a Heisenberg-scaling precision in the estimation of the parameter $\varphi$, provided that both the average number of photons in the displacement $\ND$ and in the squeezing $\NS$ scale with the total average number of photons $N=\NS+\ND$, namely $\NS= \beta N$ and $\ND = (1-\beta)N$, for any value $0<\beta\leq 1$ independent of $N$.

Moreover, it is worth noticing that the first term in equation~\eqref{eq:Fisher}, and thus in equation~\eqref{eq:FisherAsymptotic}, depends on the information encoded in the displacement of the probe, and thus it vanishes if $\Vec{\mu} = 0$, namely if the probe is a squeezed vacuum and $\ND=0$. 
The second term instead depends on the information on $\varphi$ encoded in the variance of the measurement itself: it arises only from the interaction with the squeezed photons, and vanishes if $\partial_\varphi \abs{\Sigma} = 0$, namely when $k_\mathrm{avg} = 0$ in Eq.~\eqref{eq:FisherAsymptotic}, and in particular when $\gamma_i = \pm \pi/2$, for $i=1,\dots,M$, in equation~\eqref{eq:HLConditions}, corresponding to quadratures with minimum squeezed variances, and thus locally insensible to the variations of the parameter.

Interestingly, this latter case is similar to the single squeezed vacuum and single-homodyne scenario found in the literature~\cite{Gramegna2021njp,Gramegna2021Typicality}: in fact, the second term in~\eqref{eq:FisherAsymptotic} represents a generalization of the single-homodyne Fisher information $\mathcal{F}_1(\varphi) = 8\varrho(k)(\partial_\varphi \gamma)^2 N^2$, and it can be obtained by substituting $k$ and $\partial_\varphi \gamma$ with their averages over the probabilities $P_i$, namely $k\rightarrow \sum_i P_i k_i$ and $\partial_\varphi \gamma \rightarrow \sum_i P_i \partial_\varphi \gamma_i$.  

We have then found that, also when employing multiple homodyne detections, one for each output port of the interferometer, the Heisenberg scaling precision obtained through measurements of the squeezed noise (i.e. $\partial_\varphi\Vec{\mu}=0$) is only reached when the quantum fluctuations of the observed quadratures are reduced to their quantum limit, i.e. $\abs{\Sigma} = \O(\NS^{-1})$, whilst the variations of the unknown parameter $\varphi$ still yield a visible effect on the outcomes of the measurements, i.e. $\partial_\varphi \abs{\Sigma}$ is not vanishing. 

However, at the expense of introducing a non-zero displacement in the probe, it is possible to relax the condition $\partial_\varphi \abs{\Sigma}\neq 0$, thus allowing us to choose $k_i=0$ in Eq.~\eqref{eq:DetAsymptotic}, thus effectively measuring the maximally squeezed quadratures at $\gamma_i= \pm\pi/2$. Indeed in such a case, even if the contribution to the Fisher information  associated with only the squeezed photons in Eq.~\eqref{eq:FisherAsymptotic} is vanishing, it is still possible to to reach Heisenberg scaling precision through the information on the parameter encoded in the displacement of the probe.

An important feature of this protocol, which differentiate it from its single-homodyne counterpart, is that it does \emph{not} require any adaptation of the network to the value of the unknown parameter, namely no auxiliary networks needs to be added at the input nor the output of $\hUphi$ to reach Heisenberg scaling precision. The only condition~\eqref{eq:HLConditions} can be thought as a minimum-resolution requirement on the local oscillators phases, which can thus be achieved without adding further auxiliary networks. 

However, this does not mean that the form of the network $\Uphi$ does not affect the precision of our protocol in the estimation of $\varphi$: the terms $k_\mathrm{avg}$ and $(\partial\gamma)_\mathrm{avg}$ appearing in the constant factor in the Fisher information in equation~\eqref{eq:FisherAsymptotic} depend on the transition probabilities $P_i$ and on the derivatives of the relative phases $\partial_\varphi\gamma_i$. In particular, an exceptionally poorly conceived network, e.g.\ one for which $\gamma_i$ is independent on $\varphi$ for every $i$ such that $P_i \neq 0$, can be associated with a null factor $(\partial \gamma)_\mathrm{avg}$ that sets to zero the Fisher information. In this case, adding a $\varphi$-independent auxiliary network $V$, either at the input or at the output of $\Uphi$, might modify both $P_i$ and $\gamma_i$, and thus $(\partial \gamma)_\mathrm{avg}$.

\section{Conclusions}

We have shown that performing homodyne measurements at each output channel of an arbitrary linear network encoding an unknown distributed parameter $\varphi$ to be estimated, allows us to reach Heisenberg scaling precision for a single-mode squeezed probe with no prior information on $\varphi$. 
The information on $\varphi$ is encoded both in the displacement and in the squeezing of the probe, leading to two independent contributions which can both provide Heisenberg scaling sensitivity. 
We have shown that the determinant of the covariance matrix associated with the measurement outcomes plays an important role in the enhanced sensitivity: in particular, we demonstrated that the conditions to reach Heisenberg scaling in either of the two contributions, which can be met manipulating the phases of the local oscillators, correspond to imposing that the determinant of the covariance matrix is of order $N^{-1}$ for large $N$.
Differently from protocols involving only homodyne measurements at a single channel, here there is no need for a refocusing auxiliary stage: the procedure is independent of the network and of the value of the parameter.
This allows us to safely entrust the measurement operation to an independent party without sharing any information on the structure of the network, possibly opening up a further path towards secure sensing and cryptographic quantum metrology~\cite{Huang2019,Yin2020,Shettell2021}.
On the other hand, we showed that, despite not required to achieve the Heisenberg limit, one can still employ an auxiliary stage to further enhance the estimation precision by a constant factor. 

\section{Acknowledgements}
This work was supported by the Office of Naval Research Global (N62909-18-1-2153). PF is partially supported by Istituto Nazionale di Fisica Nucleare (INFN) through the project QUANTUM, and by the Italian National Group of Mathematical Physics (GNFM-INdAM).

\appendix
\section{Joint Detection  Probability}
\label{app:Prob}
Here we will first obtain the expectation value $\Vec{\mu}$ associated with
 the homodyne measurements on the probe after the interaction with the linear network shown in~\eqref{eq:AverageVector}, then we derive the expression of the covariance matrix $\Sigma$ shown in~\eqref{eq:CovMat}, and its determinant $\abs{\Sigma}$ in~\eqref{eq:Det}.

The initial phase-space displacement  of the injected probe $\Vec{\alpha}_0 = \bra{\Psi_{\mathrm{in}}} \hat{\bm{z}} \ket{\Psi_{\mathrm{in}}}$, where 
$\hat{\bm{z}} = (\hat{x}_1, \dots, \hat{x}_M, \hat{p}_1, \dots, \hat{p}_M)$ and $\ket{\Psi_{\mathrm{in}}} = \hat{D}_1(d)\hat{S}_1(r)\vac$ is a $2M$-vector
\begin{equation}
\Vec{\alpha}_0=\begin{pmatrix}
d\\
0\\
\vdots\\
0
\end{pmatrix},
\end{equation}
with $d=\sqrt{\ND}$.
This vector is transformed by the linear network, and at the output reads
\begin{equation}
\Vec{\alpha}=R\Vec{\alpha}_0 = d\begin{pmatrix}
\sqrt{P_1}\cos\bar{\gamma}_1\\
\vdots\\
\sqrt{P_M}\cos\bar{\gamma}_M\\
\sqrt{P_1}\sin\bar{\gamma}_1\\
\vdots\\
\sqrt{P_M}\sin\bar{\gamma}_M\\
\end{pmatrix}
\label{eq:AppDisplacement}
\end{equation}
where $R$ is the $2M\times2M$ orthogonal and symplectic matrix associated with the interferometer unitary matrix $\Uphi$
\begin{equation}
R =
\begin{pmatrix}
\re{\Uphi} & -\im{\Uphi}\\
\im{\Uphi} & \re{\Uphi}
\end{pmatrix}.
\label{eq:RotationPhaseSpace}
\end{equation}
The local oscillator phases in the homodyne measurements are described by  the $2M\times 2M$ orthogonal matrix
\begin{equation}
O_{\vec{\theta}} = 
\begin{pmatrix}
\cos(\Theta) & \sin(\Theta) \\
-\sin(\Theta) & \cos(\Theta)
\end{pmatrix},
\label{eq:LocalOscillatorMatrix}
\end{equation}
with $\Theta = \diag(\vec{\theta}) = \diag(\theta_1,\dots,\theta_M)$. 
This matrix represents a clock-wise rotation in phase space for each of the $M$ channels of the network, of angles $\theta_i$ for the $i$-th mode. The mean vector $\Vec{\mu}$ in equation~\eqref{eq:AverageVector} is then given by the first $M$ elements of $O_{\vec{\theta}}\Vec{\alpha}$.

The $2M \times 2M$ symplectic covariance matrix $\Gamma_0$ of the squeezed state $\hat{S}_1(r )\ket{\mathrm{vac}}$ reads
\begin{equation}
\Gamma_0 =\dfrac{1}{2}
\begin{pmatrix}
\e^{2\mathcal{R}} & 0\\
0 & \e^{-2\mathcal{R}}
\end{pmatrix},
\end{equation}
where $\mathcal{R}$  is the $M\times M$ diagonal matrix $\mathcal{R} = \diag(r,0,\dots,0)$. 
Once again, the action of the linear network $\Uphi$ is represented by the orthogonal and symplectic matrix $R$ in equation~\eqref{eq:RotationPhaseSpace}, so that the covariance matrix of the probe at the output is
\begin{equation}
\label{eq:AppCovMat}
\Gamma = R \Gamma_0 R^T
= \begin{pmatrix}
\Sigma_x & \Sigma_{xp}\\
\Sigma_{xp}^T & \Sigma_p
\end{pmatrix},
\end{equation}
where, by direct calculation,
\begingroup
\allowdisplaybreaks
\begin{align}
\Sigma_x 
&\equiv  \dfrac{1}{2}\left[\re{\Uphi}\e^{2\mathcal{R}}\re{\Uphi^\dag} - \im{\Uphi}\e^{-2\mathcal{R}}\im{\Uphi^\dag}\right]\notag\\
&=\dfrac{1}{2}\left[\Re[\Uphi\cosh(2\mathcal{R})\Uphi^\dag]+\Re[\Uphi\sinh(2\mathcal{R})\Uphi^\mathrm{T}]\right],\\
\Sigma_p
&\equiv \dfrac{1}{2}\left[ -\im{\Uphi}\e^{2\mathcal{R}}\im{\Uphi^\dag} + \re{\Uphi}\e^{-2\mathcal{R}}\re{\Uphi^\dag}\right]\notag\\
&= \dfrac{1}{2}\left[\Re[\Uphi\cosh(2\mathcal{R})\Uphi^\dag]-\Re[\Uphi\sinh(2\mathcal{R})\Uphi^\mathrm{T}]\right],\\
\Sigma_{xp}
&\equiv \dfrac{1}{2}\left[ -\re{\Uphi}\e^{2\mathcal{R}}\im{\Uphi^\dag} - \im{\Uphi}\e^{-2\mathcal{R}}\re{\Uphi^\dag}\right]\notag\\
&=\dfrac{1}{2}\left[-\Im[\Uphi\cosh(2\mathcal{R})\Uphi^\dag]+\Im[\Uphi\sinh(2\mathcal{R})\Uphi^\mathrm{T}]\right].
\end{align}
\endgroup
The covariance matrix $\Sigma=\Sigma_x$ at the detection stage is then obtained by extracting the first $M$ rows and columns from the matrix $O_{\vec{\theta}} \Gamma O^\mathrm{T}_{\vec{\theta}}$, and thus its elements as shown in~\eqref{eq:CovMat} can be easily obtained.

To evaluate the determinant $\abs{\Sigma}$, we first notice that $\Gamma_0\equiv \mathbb{I}/2 + K_0$, where $\mathbb{I}$ is the identity matrix and $K_0 = \Gamma_0 - \mathbb{I}/2$ is a diagonal matrix of rank $2$. 
Being the rank invariant under orthogonal rotations, the same holds true for $O_{\vec{\theta}} \Gamma O^\mathrm{T}_{\vec{\theta}} \equiv \mathbb{I}/2 + K$, with $\rank(K) = 2$. 
By definition of rank, none of the sub-matrices of $K$ can have rank greater than $2$, hence we can write 
\begin{equation}
\Sigma = \mathbb{I}/2 + (\Sigma - \mathbb{I}/2) \equiv \mathbb{I}/2 + A,
\end{equation}
with $\rank(A) \leq 2$, being $A$ a submatrix of $K$. 
We can then apply the result presented in appendix~\ref{app:Theorem} to $\Sigma$ and write $\abs{\Sigma}$ as a sum of determinants of the matrices obtained replacing any number of columns of $\mathbb{I}/2$, with the respective columns of $A$
\begin{equation}
\abs{\Sigma} = \frac{1}{2^M} + \frac{1}{2^{M-1}}\sum_{i=1}^M A_{ii} + \frac{1}{2^{M-2}}\sum_{i=1}^M\sum_{j=i+1}^M A_{ii}A_{jj} - A_{ij}^2,
\end{equation}
where the first term is the determinant of $\mathbb{I}/2$, the terms in the first summation are the contributions that arise substituting the $i$-th column of $\mathbb{I}/2$ with the $i$-th column of $A$, and the terms in the last summations from substituting the $i$-th and $j$-th columns, and we also exploited the symmetry of $A$. Noticeably, since $\rank{A}\leq 2$, all the contributions involving the replacement of three or more columns of $A$ are vanishing. By direct calculation, the expression in~\eqref{eq:Det} can then be easily obtained.

\section{Fisher Information}
\label{app:Fisher}

In this Appendix we will obtain the expression for the Fisher information shown in equation~\eqref{eq:Fisher}, and the expression for the cofactor matrix in equation~\eqref{eq:CofMat}.

By plugging the probability density function in equation~\eqref{eq:pdf} into the definition of the Fisher information~\eqref{eq:FisherDef}, one easily obtain
\begin{align}
\mathcal{F}(\varphi) &= \partial_\varphi\Vec{\mu}^\mathrm{T}\Sigma^{-1}\partial_\varphi \Vec{\mu} + \frac{1}{2}\Tr[(\Sigma^{-1}\partial_\varphi \Sigma)^2]\notag\\
& = \partial_\varphi\Vec{\mu}^\mathrm{T}\Sigma^{-1}\partial_\varphi \Vec{\mu} - \frac{1}{2}\Tr[(\partial_\varphi\Sigma^{-1})(\partial_\varphi \Sigma)],
\end{align}
where we used the matrix identity $\partial_\varphi \Sigma^{-1}=-\Sigma^{-1}(\partial_\varphi \Sigma )\Sigma^{-1}$. We now can express the inverse of the covariance matrix in terms of its cofactor matrix $C$ and its determinant, namely $\Sigma^{-1} = C/\abs{\Sigma}$, where the symmetry of the covariance matrix allows us to consider directly the cofactor matrix, and not its transpose. The second term reads
\begin{align}
- \frac{1}{2}\Tr[(\partial_\varphi\Sigma^{-1})(\partial_\varphi \Sigma)] &= \frac{1}{2} \frac{\partial_\varphi \abs{\Sigma}}{\abs{\Sigma}^2}\Tr[C \partial_\varphi \Sigma] \notag \\
&\quad - \frac{1}{2\abs{\Sigma}}\Tr[(\partial_\varphi C)(\partial_\varphi \Sigma)] .
\label{eq:TempStep}
\end{align}
We  recognise in the first term of equation~\eqref{eq:TempStep}  Jacobi's formula for the derivative of the determinant, 
$\partial_\varphi \abs{\Sigma} = \Tr[C \partial_\varphi \Sigma]$,
which allows us to obtain the expression shown in equation~\eqref{eq:Fisher}
\begin{align}
\mathcal{F}(\varphi) &= \frac{1}{\abs{\Sigma}}\partial_\varphi\Vec{\mu}^\mathrm{T}C\partial_\varphi \Vec{\mu} +  \frac{1}{2}\left(\frac{\partial_\varphi\abs{\Sigma}}{\abs{\Sigma}}\right)^2 \notag\\
&\quad - \frac{1}{2\abs{\Sigma}}\Tr[(\partial_\varphi\Sigma)(\partial_\varphi C)].
\end{align}

In order to explicit the cofactor matrix $C$ in terms of the elements of $\Sigma$, and thus in terms of squeezing parameter $r$, transition probabilities $P_i$, and relative phases acquired $\gamma_i$, $i=1,\dots, M$, as displayed in~\eqref{eq:CofMat}, we first need to make some observations. 
First, the $(s,t)$-cofactor $C_{st}$, which is defined as the determinant of the $L-1\times L-1$ sub-matrix of $\Sigma$ obtained deleting the $s$-th row and $t$-th column, then multiplied by $(-1)^{s+t}$, can also be thought as the determinant  of the $L\times L$ matrix $\Sigma^{[s,t],1}$, where we denote with $X^{[s,t],n}$ the matrix obtained from the matrix $X$ replacing all the elements in the the $s$-th row and in the $p$-th column with zeros, except the element $(s,t)$ which is replaced by $n$, namely
\begin{equation}
C_{st} = \abs{\begin{pmatrix}
\Sigma_{11} & \dots & \Sigma_{1t-1} & 0 & \Sigma_{1t+1} & \dots & \Sigma_{1L}\\
 & \vdots & & \vdots & & \vdots \\
\Sigma_{s-11} & \dots & \Sigma_{s-1t-1} & 0  & \Sigma_{s-1t+1} & \dots & \Sigma_{s-1L}\\
0 & \dots & 0 & 1 & 0 & \dots & 0 \\
\Sigma_{s+11} & \dots & \Sigma_{s+1t-1} & 0  & \Sigma_{s+1t+1} & \dots & \Sigma_{s+1L}\\
 & \vdots & & \vdots & & \vdots \\
\Sigma_{L1} & \dots & \Sigma_{Lt-1} & 0 & \Sigma_{Lt+1} & \dots & \Sigma_{LL}\\
\end{pmatrix}
}.
\end{equation}
Second, as  discussed in Appendix~\ref{app:Prob}, $\Sigma = \mathbb{I}/2 + A$, where $A$ is a symmetric matrix with $\rank(A) = \rho \leq 2$. Thus, we can write the $(s,t)$-cofactor as $C_{st}=\abs{(\mathbb{I}/2)^{[s,t],0} + A^{[s,t],1}}$, and evaluate this determinant as a sum of determinants of matrices obtained swapping columns of $(\mathbb{I}/2)^{[s,t],0}$ and $A^{[s,t],1}$, as discussed in detail in Appendix~\ref{app:Theorem}. 
Noticeably, by replacing a row and a column of $A$ may increase its rank by one, so that $\rank(A^{[s,t],1})\leq 3$. It is convenient now to consider separately the simpler case $s=t$ first, and then $s\neq t$.

We notice that the matrix $(\mathbb{I}/2)^{[s,s],0}$ has a single zero eigenvalue, and thus each contribution to $C_{ss}$ is non-vanishing only if the $s$-th columns of $(\mathbb{I}/2)^{[s,s],0}$ is replaced. We thus obtain
\begin{equation}
\label{eq:appCss}
C_{ss}= \frac{1}{2^{M-1}} + \frac{1}{2^{M-2}}\sum_{\substack{i=1 \\ i\neq s}}^M A_{ii} - \frac{1}{2^{M-3}}\sum_{\substack{i=1\\i\neq s}}^M\sum_{\substack{j=i+1\\j\neq s}}^M A_{ii}A_{jj}-A_{ij}^2,
\end{equation}
which is the sum of terms obtained substituting the $s$-th, the $s$-th and $i$-th, and the $s$-th, $i$-th and $j$-th columns respectively. Noticeably, replacing more than $3$ columns yields vanishing contributions, since $\rank(A^{[s,t],1})\leq 3$. 
When $s\neq t$, $(\mathbb{I}/2)^{[s,t],0}$ has two null eigenvalues; hence, all the non-vanishing contributions must replace the $s$-th and $t$-th columns. For example, the only contribution obtained swapping the $s$-th and $t$-th columns is of the type
\begin{equation}
\frac{1}{2^{M-2}}\abs{
\begin{pmatrix}
0 & 1 \\
A_{ts} & 0
\end{pmatrix}
} = -\frac{1}{2^{M-2}} A_{st}, 
\end{equation}
where we also exploited the symmetry of $A$; the contribution obtained swapping the $s$-th, $t$-th and $i$-th columns, with $i\neq s,t$ are of the type
\begin{equation}
\frac{1}{2^{M-3}}\abs{
\begin{pmatrix}
0 & 1 & 0 \\
A_{ts} & 0 & A_{ti} \\
A_{is} & 0 & A_{ii}
\end{pmatrix}
} = \frac{1}{2^{M-3}} (A_{si}A_{ti}-A_{st}A_{ii}), 
\end{equation}
where once again, we exploited the symmetry of $A$. Replacing more than $3$ columns once again yields no contributions since $\rank(A^{[s,t],1})\leq 3$.
The final expression for $C_{st}$, $s\neq t$ thus reads
\begin{equation}
\label{eq:appCsp}
C_{st} = -\frac{1}{2^{M-2}} A_{st} + \frac{1}{2^{M-3}} \sum_{\substack{i=1\\ i\neq s,t}}^M(A_{si}A_{ti}-A_{st}A_{ii}).
\end{equation}
Replacing in~\eqref{eq:appCss} and~\eqref{eq:appCsp} the definition of $A = \Sigma - \mathbb{I}/2$, it is straightforward to obtain Eqs.~\eqref{eq:CofMat}.
\section{Asymptotics}
\label{app:Asymptotics}

In this appendix we will study the asymptotic regime of the Fisher information~\eqref{eq:Fisher},
\begin{align}
\mathcal{F}(\varphi) &= \frac{1}{\abs{\Sigma}}\partial_\varphi\Vec{\mu}^\mathrm{T}C\partial_\varphi \Vec{\mu} +  \frac{1}{2}\left(\frac{\partial_\varphi\abs{\Sigma}}{\abs{\Sigma}}\right)^2 \notag\\
&\quad - \frac{1}{2\abs{\Sigma}}\Tr[(\partial_\varphi\Sigma)(\partial_\varphi C)],
\end{align}
for large $N = \NS + \ND = \sinh^2 r + d^2$. We will show that the only conditions needed to reach Heisenberg scaling are the ones shown in~\eqref{eq:HLConditions}, and that in this case the asymptotic expression for the Fisher information is~\eqref{eq:FisherAsymptotic}.

First, from the explicit expressions of $\Sigma$ in~\eqref{eq:CovMat} and $C$ in~\eqref{eq:CofMat}, we notice that each of their matrix elements are at most of order of $\NS$, since $\sinh r = \sqrt{\NS}$, the single-photon probabilities $P_i$ are independent of $\NS$,  and the cosine and sine functions are limited. 

The same holds true for their derivatives: in particular $\partial_\varphi P_i$ and $\partial_\varphi \gamma_i = \partial_\varphi \bar{\gamma}_i$ do not depend on $\NS$, where $\gamma_i = \bar{\gamma}_i-\theta_i$ is the phase acquired by the signal through the $i$-th output port relatively to the local-oscillators reference phases. 
Moreover, similar considerations can be applied for $\Vec{\mu} = \O(\sqrt{\ND})$, and specifically for its derivative $\partial_\varphi \Vec{\mu}$.
This implies that, in order to reach a Heisenberg-scaling sensitivity, namely a scaling of order of $N^2$ in the Fisher information, the determinant $\abs{\Sigma}$ cannot be of any order higher than $N^0$. 

We thus first focus our attention on $\abs{\Sigma}$
\begin{align}
\abs{\Sigma} = \frac{1}{2^M} &+ \frac{\sinh(r)}{2^{M-1}}\sum\limits_{i=1}^M P_i (\sinh(r) + \cos(2\gamma_i)\cosh(r)) \notag\\
&-\frac{\sinh^2(r)}{2^{M-2}}\sum\limits_{i = 1}^M \sum\limits_{j = i+1}^M P_i P_j \sin^2(\gamma_i-\gamma_j),
\label{eq:AppDet}
\end{align}
shown in equation~\eqref{eq:Det}.
In particular we will suppose that, for large $N$, $\gamma_i$ tend to finite values $\gamma_{0i}$, namely that $\gamma_i = \gamma_{0i} + k_i N^{-\alpha}$, with $k_i\in\R$ of order 1 and $\alpha > 0$, since if $\gamma_i$ were to grow with $N$ (i.e. $\alpha<0$), it would give rise to an oscillating asymptotic behaviour to $\abs{\Sigma}$.

By expanding the squeezing parameter $r$ in powers of $\NS$ in~\eqref{eq:AppDet}, we obtain
\begin{align}
\abs{\Sigma} = D_1 \NS + D_2 + D_3\frac{1}{\NS} + \O\left(\frac{1}{\NS^2}\right),
\label{eq:appDet}
\end{align}
where
\begin{align}
D_1 &= \frac{1}{2^{M-1}}\biggl(1+\sum_{i=1}^M P_i\cos(2\gamma_{i}) \biggr) \notag\\
&- \frac{1}{2^{M-2}}\biggl(\sum_{i=1}^M\sum_{j=i+1}^M P_i P_j\sin(\gamma_i-\gamma_j)^2\biggr),\\
D_2 &= \frac{1}{2^M}\biggl(1 + \sum_{i=1}^M P_i \cos(2\gamma_i)\biggr),\\
D_3 &= -\frac{1}{2^{M+2}}\sum_{i=1}^M P_i\cos(2\gamma_i).
\end{align}
In order to cancel the scaling with $\NS$, $D_1$ must be equal to, or tend to zero. 

After some trigonometry, and exploiting the passivity of the linear network $\Uphi$ which sets $\sum_i P_i = 1$, we can rewrite
\begin{equation}
D_1 = \frac{1}{2^{M-2}}\biggl( \Bigl(\sum_{i=1}^M P_i \cos^2\gamma_i\Bigr)^2 + \Bigl(\sum_{i=1}^M P_i \sin\gamma_i\cos\gamma_i \Bigr)^2\biggr),
\end{equation}
which tends to zero iff $\gamma_{0i} = \pi/2 + n\pi$, with $n\in\mathbb{Z}$. 

In particular, the scaling of $\abs{\Sigma}$ for large $\NS$ will be of order $\NS^0$ or lower only if $\gamma_i = \pi/2 + n\pi + k_i \NS^{-\alpha}$, with $\alpha \geq 1/2$.
To see that, we notice that, for $\gamma_{0i}=\pi/2$, $D_1$ and $D_2$ scale with $\NS^{-2\alpha}$, while $D_3$ scale with $\NS^0$.
Thus, $\abs{\Sigma}$ scales with $\NS^{1-2\alpha}$ for $\alpha \leq 1$ (and in particular with $\NS^0$ for $\alpha=1/2$), and with $\NS^{-1}$ for $\alpha > 1$. 
Noticeably, also $D_2$ tends to zero iff $\gamma_{0i} = \pi/2 + n\pi$. 

We now study the asymptotics of the numerators appearing in the Fisher information, when the condition $\gamma_i = \pi/2 + k_i \NS^{-\alpha}$, with $\alpha \geq 1/2$, is true. 
We first obtain the derivative of $\Sigma$ from equation~\eqref{eq:CovMat}, and substitute $\gamma_i = \pi/2 + k_i/\NS^\alpha$, with $\alpha\geq 1/2$,
\begin{align}
\partial_\varphi \Sigma_{ij}=& -\frac{1}{2}\partial_\varphi(\sqrt{P_i P_j}) +\sqrt{P_i P_j}\Big(\partial_\varphi(\gamma_i+\gamma_j)(k_i+k_j)\notag\\
&-\partial_\varphi(\gamma_i-\gamma_j)(k_i-k_j)\Big)\NS^{1-\alpha}
\nonumber\\
& +\O(\NS^{1-2\alpha})+\O(\NS^{-1}),
\label{eq:dSigma}
\end{align}
and we notice that it scales at most with $N^{1-\alpha}$ for $1/2 \leq  \alpha < 1$, and at most with $N^0$ for $\alpha\geq 1$. 
We then analyse the auxiliary term~\eqref{eq:STerms}
\begin{align}
S_{sti} & = \sinh^2(r)\sqrt{P_s P_t}P_i \sin(\gamma_s-\gamma_i)\sin(\gamma_t-\gamma_i) 
\notag\\
&= \O(\NS^{1-2\alpha}),
\label{eq:S}
\end{align}
of which we evaluate the derivative when $\gamma_i = \pi/2 + k_i/\NS^\alpha$, with $\alpha\geq 1/2$, $i=1,\dots,M$
\begin{align}
\partial_\varphi S_{sti} &= \sqrt{P_sP_t}P_i\Big((\partial_\varphi(\gamma_s-\gamma_i))(k_t-k_i)\notag\\
&+(\partial_\varphi(\gamma_t-\gamma_i))(k_s-k_i)\Big)\NS^{1-\alpha} + \O(N^{1-2\alpha}),
\label{eq:dS}
\end{align}
which scales at most with $\NS^{1-\alpha}$ for large $\NS$. 

Moreover, the covariance matrix $\Sigma$ in~\eqref{eq:CovMat} asymptotically reads
\begin{equation}
\Sigma_{ij} = \frac{\delta_{ij}-\sqrt{P_i P_j}}{2} + \O(\NS^{1-2\alpha}) + \O(\NS^{-1}),
\label{eq:CovMatAsymptotics}
\end{equation}
and thus, inserting~\eqref{eq:S} and~\eqref{eq:CovMatAsymptotics} in the cofactor matrix~\eqref{eq:CofMat}, we obtain
\begin{equation}
C_{st} = \frac{1}{2^{M-1}}\sqrt{P_s P_t} + \O(\NS^{1-2\alpha}) + \O(\NS^{-1}).
\label{eq:CofactorAsymptotic}
\end{equation}
Finally, we can write
\begin{align}
\partial_\varphi \abs{\Sigma} &= \frac{1}{2^{M-1}}\sum_{i=1}^M \partial_\varphi \Sigma_{ii} - \frac{1}{2^{M-2}}\sum_{i=1}^M\sum_{j=i+1}^M \partial_\varphi S_{iij}\label{eq:dDet},\\
\partial_\varphi C_{ss} &=\frac{1}{2^{M-2}} \sum_{\substack{i=1\\i\neq s}}^M\partial_\varphi \Sigma_{ii} - \frac{1}{2^{M-3}}\sum_{\substack{i=1 \\ i\neq s}}^M\sum_{\substack{j=i+1\\j\neq s}}^M \partial_\varphi S_{iij},\\
\partial_\varphi C_{st} &= -\frac{1}{2^{M-2}}\partial_\varphi \Sigma_{st} + \frac{1}{2^{M-3}}\sum_{\substack{i=1\\i \neq s,t}}^M \partial_\varphi S_{sti}.
\end{align}
It is easy to see that $\partial_\varphi \abs{\Sigma}$ scales at most with $\NS^{1-\alpha}$, since $\sum_i^M \partial_\varphi P_i = 0$, while the derivatives of the elements of $C$ scale at most with $\NS^{1-\alpha}$ for $1/2\leq\alpha<1$, and at most with $\NS^0$ for $\alpha\geq 1$. 

Our last step is to evaluate the asymptotics for $\partial_\varphi \Vec{\mu}$, which can be easily evaluated differentiating equation~\eqref{eq:AverageVector}
\begin{align}
\partial_\varphi\mu_i &= \sqrt{\ND}\left(\frac{\partial_\varphi P_i}{2\sqrt{P_i}}\cos\gamma_i - \sqrt{P_i}\partial_\varphi\gamma_i \sin\gamma_i\right)\notag \\
&= -\sqrt{\ND P_i}\partial_\varphi\gamma_i + \O(\sqrt{\ND} \NS^{-\alpha}).
\label{eq:MuDerivative}
\end{align}

Now we can finally draw our conclusions, and obtain the scaling of the Fisher information~\eqref{eq:Fisher} by putting together all the asymptotic regimes found. First, we notice that, independently of $\alpha\geq 1/2$, the last term in equation~\eqref{eq:Fisher} always scales with $\NS$, hence only reaching shot-noise precision. 
The only terms which allow the Heisenberg-scaling are then the first two: in fact, the first term scales with $\ND\NS^{2\alpha-1}$ for $1/2\leq\alpha\leq 1$, and with $\ND\NS$ for $\alpha > 1$, and thus reaching Heisenberg scaling for $\alpha\geq 1$ (condition $\alpha>1$ includes the case $\gamma_i=\pi/2$), while the second term reaches sub-shot noise scaling $\NS^{2-2\abs{\alpha-1}}$ for $1/2 < \alpha < 3/2$, with Heisenberg scaling for $\alpha=1$. 

This shows that the only condition to reach Heisenberg-scaling is that 
\begin{equation}
\gamma_i = \pi/2 + \O(\NS^{-1}),
\label{eq:HLConditionApp}
\end{equation}
for $i=1,\dots,M$, as shown in~\eqref{eq:HLConditions}, or equivalently that asymptotically $\gamma_i \simeq \pi/2 + k_i/\NS$ with $k_i\in\R$ of order 1, as well as that the only relevant terms under this condition are the first two in expression~\eqref{eq:Fisher}, and that only for $\alpha = 1$ the first term is non-vanishing.

We can now finally prove~\eqref{eq:FisherAsymptotic}. Substituting condition~\eqref{eq:HLConditionApp} inside Eqs.~\eqref{eq:dSigma} and~\eqref{eq:dS}, we obtain from~\eqref{eq:dDet} the asymptotics
\begin{align}
\partial_\varphi \abs{\Sigma} &\simeq \frac{1}{2^{M-1}}\sum_{i=1}^M 4P_i k_i (\partial_\varphi \gamma_i) \notag \\
&-\frac{1}{2^{M-2}}\sum_{i=1}^M\sum_{j=i+1}^M 2P_iP_j(k_i-k_j)(\partial_\varphi\gamma_i - \partial_\varphi \gamma_j) \notag\\
&\simeq\frac{1}{2^{M-3}}\sum_{i=1}^M  P_i k_i \sum_{j=1}^M P_j \partial_\varphi\gamma_j
\label{eq:DerDetAsymptoticApp}
\end{align}
where we once again exploited the passivity of the interferometer, so that $\sum_{i=1}^M P_i = 1$, while we obtain from~\eqref{eq:appDet}
\begin{equation}
\abs{\Sigma} \simeq \frac{1}{2^{M-2} N} \biggl(\Bigl(\sum_{i=1}^M P_i k_i\Bigr)^2 + \frac{1}{16}\biggr),
\label{eq:DetAsymptoticApp}
\end{equation}
which is the expression shown in~\eqref{eq:DetAsymptotic}. Lastly, from equation~\eqref{eq:MuDerivative} when conditions~\eqref{eq:HLConditionApp} hold, we obtain the asymptotics
\begin{equation}
\partial_\varphi\mu_i\simeq-\sqrt{\ND P_i}\partial_\varphi \gamma_i.
\label{eq:DerMuAsymptoticApp}
\end{equation} 
Eqs.~\eqref{eq:DerDetAsymptoticApp},~\eqref{eq:DetAsymptoticApp} and~\eqref{eq:DerMuAsymptoticApp} yield the asymptotic expression for the Fisher information shown in~\eqref{eq:FisherAsymptotic}.

\section{Maximum-Likelihood Estimator}
\label{app:MLE}

We will now obtain the implicit equation that defines the Maximum-Likelihood Estimator $\tilde{\varphi}_\mathrm{MLE}$, which saturates the Heisenberg-scaling Cramér–Rao bound~\eqref{eq:CRb}, with the Fisher information given in equation~\eqref{eq:FisherAsymptotic}, in the asymptotic regime of large $\nu$. 

Let us then imagine that, after $\nu$ iterations, we collect $\nu$ sets of outcomes $\Vec{x}_i$, $i=1,\dots,\nu$ from the $M$ homodyne measurements at the output of the linear network $\hUphi$. 
The likelihood that these outcomes are observed for a given value $\varphi$ of the unknown parameter is given by
\begin{equation}
\mathcal{L}(\varphi | \Vec{x}_1,\dots,\Vec{x}_\nu) = \prod_{j=1}^\nu p( \Vec{x}_j | \varphi),
\end{equation}
where $p( \Vec{x}_j | \varphi)$ is the probability density function given in~\eqref{eq:pdf}. 
The Maximum-Likelihood estimator $\tilde{\varphi}\equiv\tilde{\varphi}(\Vec{x}_1,\dots,\Vec{x}_\nu)$ is defined as the value of $\varphi$ which maximizes the likelihood $\mathcal{L}$ that the outcomes $\Vec{x}_1,\dots,\Vec{x}_\nu$ are observed, and it is usually found by maximising the log-likelihood function
\begin{align}
0&=\partial_\varphi\log\mathcal{L}(\varphi | \Vec{x}_1,\dots,\Vec{x}_\nu)\Big\vert_{\varphi=\tilde{\varphi}_\mathrm{MLE}} \notag \\
&= \partial_\varphi\sum_{j=1}^\nu \log p(\Vec{x}_j | \varphi)\Big\vert_{\varphi=\tilde{\varphi}_\mathrm{MLE}} \notag \\
&= \biggl[-\frac{\nu}{2}\partial_\varphi\log(\abs{\Sigma})
\notag\\
&\quad\quad-\frac{1}{2}\partial_\varphi\sum_{j=1}^\nu(\vx_j-\Vec{\mu})^\mathrm{T}\Sigma^{-1}(\vx_j-\Vec{\mu})\biggr]_{\varphi=\tilde{\varphi}_\mathrm{MLE}}\notag\\
&= \biggl[-\frac{\nu}{2}\Tr[\Sigma^{-1}\partial_\varphi\Sigma]
\notag\\
&\quad\quad-\frac{1}{2}\partial_\varphi\sum_{j=1}^\nu\Tr[\Sigma^{-1}(\vx_j-\Vec{\mu})(\vx_j-\Vec{\mu})^\mathrm{T}]\biggr]_{\varphi=\tilde{\varphi}_\mathrm{MLE}}\notag\\
&= \frac{1}{2}\Tr\biggl[\partial_\varphi\Sigma^{-1}\Bigl(\nu\Sigma-\sum_{j=1}^\nu(\vx_j-\Vec{\mu})(\vx_j-\Vec{\mu})^\mathrm{T}\Bigr)\biggr]_{\varphi=\tilde{\varphi}_\mathrm{MLE}}\notag\\
&\quad +\biggl[(\partial_\varphi\Vec{\mu})^\mathrm{T}\Sigma^{-1}\Bigl(\nu\Vec{\mu}-\sum_{j=1}^\nu\vx_j\Bigr)\biggr]_{\varphi=\tilde{\varphi}_\mathrm{MLE}},
\label{eq:MLEDef}
\end{align}
where we exploited  Jacobi's formula for the derivative of the determinant of a matrix, the identity $\Tr(\Sigma^{-1}\partial_\varphi \Sigma)= -\Tr(\partial_\varphi\Sigma^{-1} \Sigma)$,
and the symmetry of $\Sigma$. Once inserting the expressions for the covariance matrix $\Sigma$ from~\eqref{eq:CovMat} and for the mean vector $\Vec{\mu}$ from~\eqref{eq:AverageVector} into the previous equation, one is able to obtain with numerical methods the Maximum-Likelihood estimator as the $\varphi$ that solves~\eqref{eq:MLEDef}.

Equation~\eqref{eq:MLEDef} largely simplifies when the probe is a squeezed-vacuum state ---i.e. $\Vec{\mu}=0$--- or when the maximally-squeezed quadratures are measured ---i.e. $\partial_\varphi \abs{\Sigma} = 0$. In the first case, it becomes
\begin{equation}
0=\frac{1}{2}\Tr\biggl[\partial_\varphi\Sigma^{-1}\Bigl(\Sigma-\frac{1}{\nu}\sum_{j=1}^\nu \vx_j\vx_j^\mathrm{T}\Bigr)\biggr]_{\varphi=\tilde{\varphi}_\mathrm{MLE}},
\label{eq:MLENoMu}
\end{equation} 
where it is possible to recognise the usual mean squared error estimator $\tilde{\Sigma} = \sum_{j=1}^\nu \vx_j\vx_j^\mathrm{T}/\nu$ for the variance matrix $\Sigma$, so that the solution of equation~\eqref{eq:MLENoMu} can be seen as the value of $\varphi$ that sets equal to zero a weighted mean of the elements of covariance matrix estimator. In the latter case, equation~\eqref{eq:MLEDef} becomes
\begin{equation}
0=\biggl[(\partial_\varphi\Vec{\mu})^\mathrm{T}\Sigma^{-1}\Bigl(\Vec{\mu}-\frac{1}{\nu}\sum_{j=1}^\nu\vx_j\Bigr)\biggr]_{\varphi=\tilde{\varphi}_\mathrm{MLE}},
\end{equation}
which can be seen as a weighted mean of the estimators $\tilde{\Vec{\mu}}=\sum_{j=1}^\nu \Vec{x}_j /\nu$ of the mean $\Vec{\mu}$.
\section{Useful formulas for the determinant of a sum of two matrices}
\label{app:Theorem}
Let us consider an $L\times L$ matrix $Z$ which can be written as $Z = D + W$, where $D = \diag(d_1,\dots,d_L)$ is a real diagonal matrix, and $\rank(W) = \rho \leq L$. In this appendix, we will show a way to write the determinant $\abs{Z}$ in terms of the elements of $W$, which is convenient for our purposes.

We exploit the identity~\cite{Xu1996}
\begin{equation}
\abs{Z} = \abs{D + W} = \sum_{\alpha=0}^L \Theta_\alpha(D,W),
\end{equation}
where $\Theta_\alpha (X,Y)$ is the sum of the determinants of the matrices obtained by replacing any set of $\alpha$ columns (rows) of $X$ with the corresponding $\alpha$ columns (rows) of $Y$. 
Since $\rank(W) = \rho$, the rank of any $\alpha\times\alpha$ sub-matrix of $W$ is zero if $\alpha>\rho$, so we can write
\begin{equation}
\abs{Z} = \sum_{\alpha=0}^\rho \Theta_\alpha(D,W).
\end{equation}

Let us make explicit the first, easier terms of the summation with the purpose to grasp the gist of this expression. 
For $\alpha = 0$, no columns are replaced from $D$, hence $\Theta_0(D,W) = \abs{D} = \prod_k d_k$. 
We notice that, if at least one of the eigenvalues of $D$ is zero, this term vanishes. For $\alpha = 1$, $\Theta_1(D,W)$ is the sum of determinants of matrices of the form
\begin{equation}
\begin{pmatrix}
d_1 & 0 & \dots & 0 & W_{1i} & 0 & \dots & 0 & 0 \\
0 & d_2 & \dots & 0 & W_{2i} & 0 & \dots & 0 & 0 \\
  &   & \ddots &  & \vdots & & \vdots & \\
0 & 0 & \dots & d_{i-1} & W_{i-1i} & 0 & \dots & 0 & 0 \\
0 & 0 & \dots & 0 & W_{ii} & 0 & \dots & 0 & 0\\
0 & 0 & \dots & 0 & W_{i+1i} & d_{i+1} & \dots & 0 & 0\\
  &   & \vdots &  & \vdots & & \ddots & \\
0 & 0 & \dots & 0 & W_{L-1i} & 0 & \dots & d_{L-1} & 0\\
0 & 0 & \dots & 0 & W_{Li} & 0 & \dots & 0 & d_{L}\\
\end{pmatrix}
\end{equation}
with $i=1,\dots ,L$. Due to the structure of this matrices, their determinant is straightforward and reduces to $W_{ii}\times\prod_{k\neq i} d_k$. This means that, in general, $\Theta_1(D,W)=\sum_i W_{ii}\times\prod_{k\neq i} d_k$. 
A key observation to make is that if two or more eigenvalues of $D$ are zero, all these determinants are zero, and thus $\Theta_1(D,W)=0$; 
if instead a single eigenvalue is zero, say $d_j=0$, with $j=1,\dots,L$, then only one of these determinants is non-vanishing, i.e. the determinant of the matrix obtained by replacing the $j$-th column of $D$. 
In this case then we have $\Theta_1(D,W)=W_{jj}\times\prod_{k\neq j} d_k$. 

Let us now consider lastly the case $\alpha = 2$. The matrices whose determinants contribute to $\Theta_2(D,W)$ are of the form
\begin{widetext}
\begin{equation}
\begin{pmatrix}
d_1 & \dots & 0 & W_{1i} & 0 & \dots & 0 & W_{1i'} & 0 & \dots & 0\\
 & \ddots & & \vdots & & \vdots & & \vdots & & \vdots & \\
0 & \dots & d_{i-1} & W_{i-1i} & 0 & \dots & 0 & W_{i-1 i'} & 0 & \dots & 0 \\
0 & \dots & 0 & W_{ii} & 0 & \dots & 0 & W_{ii'} & 0 & \dots & 0 \\
0 & \dots & 0 & W_{i+1i} & d_{i+1} & \dots & 0 & W_{i+1i'} & 0 & \dots & 0\\
 & \vdots & & \vdots & & \ddots & &\vdots & &\vdots & \\
0 & \dots & 0 & W_{i'-1i} & 0 & \dots & d_{i'-1} & W_{i'-1 i'} & 0 & \dots & 0\\
0 & \dots & 0 & W_{i'i} & 0 & \dots & 0 & W_{i'i'} & 0 & \dots & 0\\
0 & \dots & 0 & W_{i'+1 i} & 0 & \dots & 0 & W_{i'+1 i'} & d_{i'+1} & \dots & 0\\
 & \vdots & & \vdots & & \vdots & & \vdots & & \ddots & \\
0 & \dots & 0 & W_{Li} & 0 & \dots & 0 & W_{Li'} & 0 & \dots & d_L

\end{pmatrix},
\end{equation}
\end{widetext}
with $i<i'=1,\dots,L$. 
Once again, the determinants of these type of matrices are easy to be evaluated and read $|W^{(i,i')}|\,\prod_{k\neq i,i'}d_k$, where 
\begin{equation}
W^{(i,i')} = \begin{pmatrix}
W_{ii} & W_{ii'}\\
W_{i'i} & W_{i'i'}
\end{pmatrix}.
\end{equation}
We notice that $|W^{(i,i')}| = |W^{(i',i)}|$. 
Thus, in general, $\Theta_2(D,W) = \sum_i\sum_{j>1} |W^{(i,i')}|\,\prod_{k\neq i,i'}d_k$. 
Once again, key observations can be made: if $D$ has at least three null eigenvalues, then $\Theta_2(D,W)$ is vanishing. 
If only two eigenvalues are zero, e.g. $d_j=d_{j'}=0$, then there is only one contribution to $\Theta_2(D,W)$, given by the matrix obtained substituting the $j$-th and $j'$-th columns of $D$, and in this case we have $\Theta_2(D,W) = |W^{(j,j')}|\,\prod_{k\neq j,j'}d_k$. 
If only one eigenvalue is zero, namely $d_j=0$, then $\theta_2(D,W)$ is given by the sum of all the determinants of the matrices where the $j$-th column has been replaced, namely $\Theta_2(D,W) = \sum_{i\neq j}|W^{(i,j)}|\,\prod_{k\neq i,j}d_k$.

Similarly, it is possible to extend these considerations to every value of $\alpha$, and finally obtain the compact form
\begin{equation}
\abs{Z} = \sum_{\alpha=0}^\rho \sum_{\gamma\in\mathcal{C}^L_\alpha} |W^{(\gamma)}|\,\prod_{k\notin\gamma}d_k,
\end{equation}
where $\mathcal{C}^L_\alpha$ is the set of all the combinations of $\alpha$ items in a set of $L$, and $W^{(\gamma)}$ denotes the $\alpha\times\alpha$ sub-matrix of $W$ obtained by selecting the rows and columns with indices $\gamma_1,\dots,\gamma_\alpha$.

\nocite{*}
\bibliography{references}

\end{document}